\documentclass[journal]{IEEEtran}
\pdfoutput=1
\ifCLASSINFOpdf
\else
   \usepackage[dvips]{graphicx}
\fi
\usepackage{url}

\usepackage{subfigure}
\usepackage{stfloats}
\usepackage{graphicx}
\usepackage{booktabs}
\usepackage{amsmath}
\usepackage{amssymb}
\usepackage{array, caption, threeparttable}
\usepackage[font=small,labelfont=bf,labelsep=none]{caption}

\usepackage{booktabs} 
\usepackage{array}
\usepackage{multirow}
\makeatletter
\newif\if@restonecol
\makeatother

\usepackage[linesnumbered,ruled,vlined]{algorithm2e}%[ruled,vlined]{
\usepackage{algpseudocode}
\begin{document}

\title{Wake Word Detection Model Based on Res2Net}

\author{Qiuchen Yu, Ruohua Zhou}

\markboth{Journal of \LaTeX\ Class Files, Vol. X, No. X, August X}
{Shell \MakeLowercase{\textit{et al.}}: Bare Demo of IEEEtran.cls for IEEE Journals}
\maketitle

\begin{abstract}

This letter proposes a new wake word detection system based on Res2Net. As a variant of ResNet, Res2Net was first applied to objection detection. Res2Net realizes multiple feature scales by increasing possible receptive fields. This multiple scaling mechanism significantly improves the detection ability of wake words with different durations. Compared with the ResNet-based model, Res2Net also significantly reduces the model size and is more suitable for detecting wake words. The proposed system can determine the positions of wake words from the audio stream without any additional assistance. The proposed method is verified on the Mobvoi dataset containing two wake words. At a false alarm rate of 0.5 per hour, the system reduced the false rejection of the two wake words by more than 12\% over prior works.
\end{abstract}


\IEEEpeerreviewmaketitle

\section{Introduction}

\IEEEPARstart{W}{ake} word detection (WWD) is to detect a predefined keyword from streaming audio or prerecorded audio utterance. It can be regarded as a special keyword spotting task and has been widely used in smart devices to facilitate the interaction between people and virtual assistants. For voice-controlled devices such as Google Home and Amazon Echo, their WWD systems usually work continuously in the background. People can wake up and interact with these devices by saying a predefined word "Alexa" or "okay Google" \cite{keyword_search}. This requires the WWD system to have a small power consumption and latency.

Large vocabulary continuous speech recognition (LVCSR) \cite{lvcsr1,lvcsr2,lvcsr3} is a classic solution to keyword detection. Its advantage is that the keywords can be changed at any time, but it requires great computing resources. LVCSR-based systems have been designed to detect audio content in large datasets, but they are not suitable for small-footprint applications such as WWD. There are two dominant categories of WWD methods: keyword/filler Hidden Markov Models (HMM) \cite{hmm1, hmm2, hmm3}, and pure neural models \cite{pure_nn1, pure_nn2, pure_nn3}. Although the HMM approach is a classic technique, it still has strong competitiveness today. HMMs are established by using keyword and non-keyword audio segments. All valid phone sequences from the keyword are modeled by an HMM, and non-keyword speech segments are absorbed by a filler HMM. Then, Viterbi decoding is adopted to obtain the best path in the decoding graph, where the computational cost depends on the specific HMM topology. The Gaussian mixture model (GMM) was the first choice for modeling observed acoustic features. With the continuous development of deep learning, deep neural networks (DNNs) have now become an alternative for GMMs because of their better performance \cite{hmm_dnn4, hmm_dnn5, hmm_dnn6}.
 
The pure neural network approach was first proposed by Google as a small-footprint approach \cite{pure_nn1}, and it has attracted more and more attention. This approach is trained for a keyword. When the confidence score of a posterior handing algorithm is larger than the threshold, a keyword is detected \cite{pure_nn6, pure_nn7}. Meanwhile, since the size of the neural network can be controlled and there is no decoding graph, this approach can be executed on hardware platforms with a small footprint and low latency. Subsequently, feed-forward DNNs were replaced by stronger networks, such as the convolutional neural network (CNN) \cite{pure_nn10} and the recurrent neural network (RNN) \cite{pure_nn12}, and further performance improvement was obtained. 

It is necessary to activate a WWD system once the wake word appears in an audio stream. The family of RNNs \cite{hochreiter1997long, cho2014learning} may suffer from state saturation while facing a continuous input stream, increasing computational cost and detection latency even with GPUs it cannot be parallelized in the case of chunk-streaming due to its sequential computations. Though RNNs are more suitable for modeling timing signals, considering the short duration of wake words, the words can be modeled by stacking multiple CNN layers \cite{pure_nn16}. In the CNN-based WWD systems, a convolution kernel is repeatedly used by sliding it over time or frequencies, thus covering a small and fixed range of frames. Meanwhile, the receptive field is expanded by stacking multiple convolution layers, even though each kernel can only capture local patterns. The stacking of more layers can obtain larger ranges of frames than fewer layers, which helps to capture more global patterns.

The CNN-based system requires fixed sizes as input. Many approaches use non-overlapping segments that are long enough (e.g., 1 second) as input features \cite{pure_nn1, arik2017convolutional, bai2019time}. Voice activity detection is sometimes used to reduce computation by only running the WWD algorithm when the voice is present. It only detects the local feature of the wake word at a time when the duration of the wake word is longer than the default length. If the wake word is too short, they will be affected by non-wake word utterances when the global feature of the wake word is detected. Accurate audio of the wake word can provide valuable information for the WWD system.

In this letter, we propose a new approach to extract accurate wake word features for the WWD system. Specifically, we first determine the wake word regions by local features of wake words and then yield global features of the same size from wake word regions with different lengths. This not only solves the above problem but also jointly detects the local and global features of wake words, which improves the WWD system performance. To further improve the performance, this letter adapts Res2Net \cite{res2net1} whose multiple scaling mechanism is very suitable for detecting wake words with different lengths. By modifying the model structure, we reduce the parameters of the model without affecting the performance as much as possible. The contributions of this letter can be summarized as: 1) We adopt a new approach to provide the WWD system with more accurate input than before, and can process streaming audio. 2) With smaller model parameters, our proposed WWD system achieves better performance than the latest system.

\section{Res2Net-\scriptsize{BASED} \normalsize{WWD}}

\begin{figure*}[thbp]
    \centering
    \includegraphics[width=\textwidth]{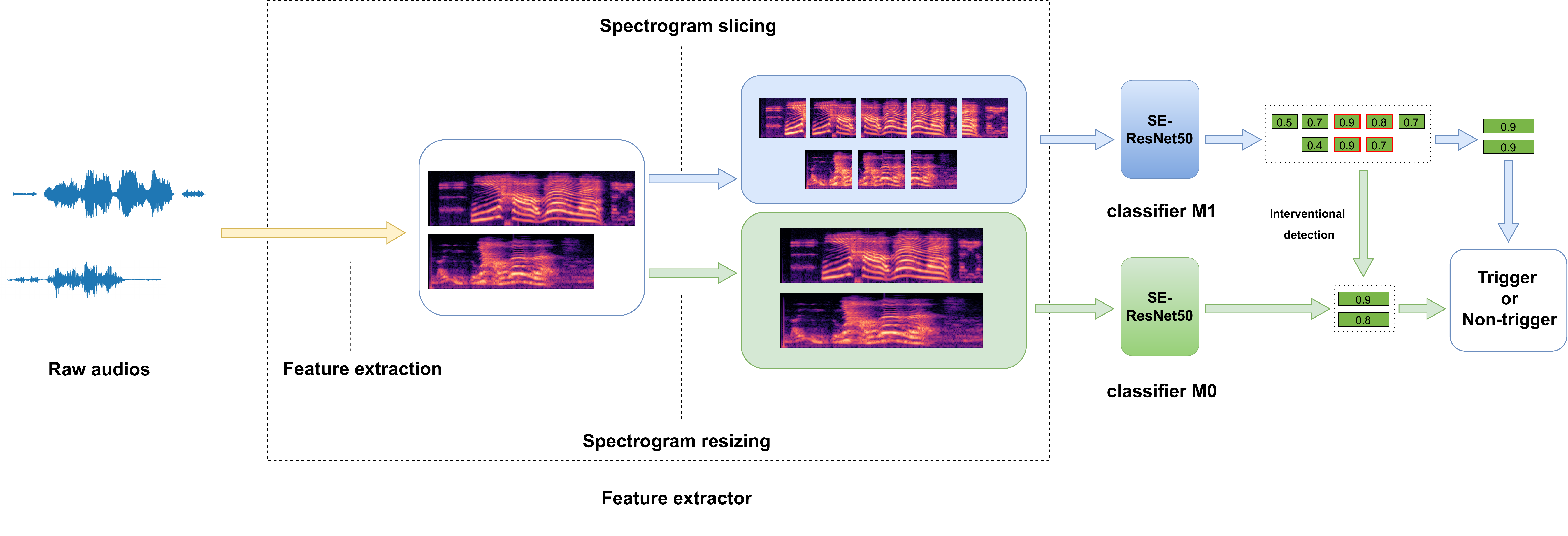}
    \caption{Our proposed WWD system. The figure shows that two types of features will be obtained after processing each audio. For classifier M0, spectrograms of the same length are used as the training input. The classifier M1 takes the sliced sub-spectrograms as input. The two classifiers output classification scores for each spectrogram. Whether the final wake word is triggered is determined by the average classification score of the two classifiers.}
\label{fig:yu1}
\end{figure*}

As demonstrated in Fig. \ref{fig:yu1}, this letter proposes a convolutional network-based WWD system. There are two modules in the system: the feature extractor, and the classifier. 

\subsection{The feature extractor}
The feature extractor takes the raw audio as input and outputs two types of processed speech spectrogram features. This module uses two processing methods: spectrogram resizing and spectrogram slicing.

\subsubsection{Spectrogram Resizing}

The length or duration of wake words is greatly affected by the speaker. Since the spectrogram of the same speech has a similar structure, this letter uses CNN to extract wake word information from the spectrogram. How to use CNN to classify spectrograms of different sizes needs full consideration. In this letter, an interpolation-based image resizing technique is adopted to obtain spectrograms of the same size. Convoluting an image with a tiny kernel and weight coefficients is a common interpolation technique \cite{sharan2019time}. In both x and y dimensions, bilinear interpolation can be regarded as an extension of linear interpolation as:

\begin{equation}
R_{BL}(x,y)=a_{0}+a_{1}x+a_{2}y+a_{3}xy
\label{bilinear interpolation}
\end{equation}

where the values of $a_{0}$, $a_{1}$, $a_{2}$, and $a_{3}$ are determined by the four nearest neighbors of $(x,y)$. The bilinear interpolation kernel can be given as

\begin{equation}
k(x)=\begin{cases}
1-|x| & |x|<1 \\
0 & \textit{otherwise} \\
\end{cases}
\label{bilinear interpolation kernel}
\end{equation}

From training spectrograms of the same size, the classifier \textit{M0} can learn the global pattern of wake words.

\subsubsection{Spectrogram Slicing}
\begin{figure}[h] \centering
    \includegraphics[width=\columnwidth]{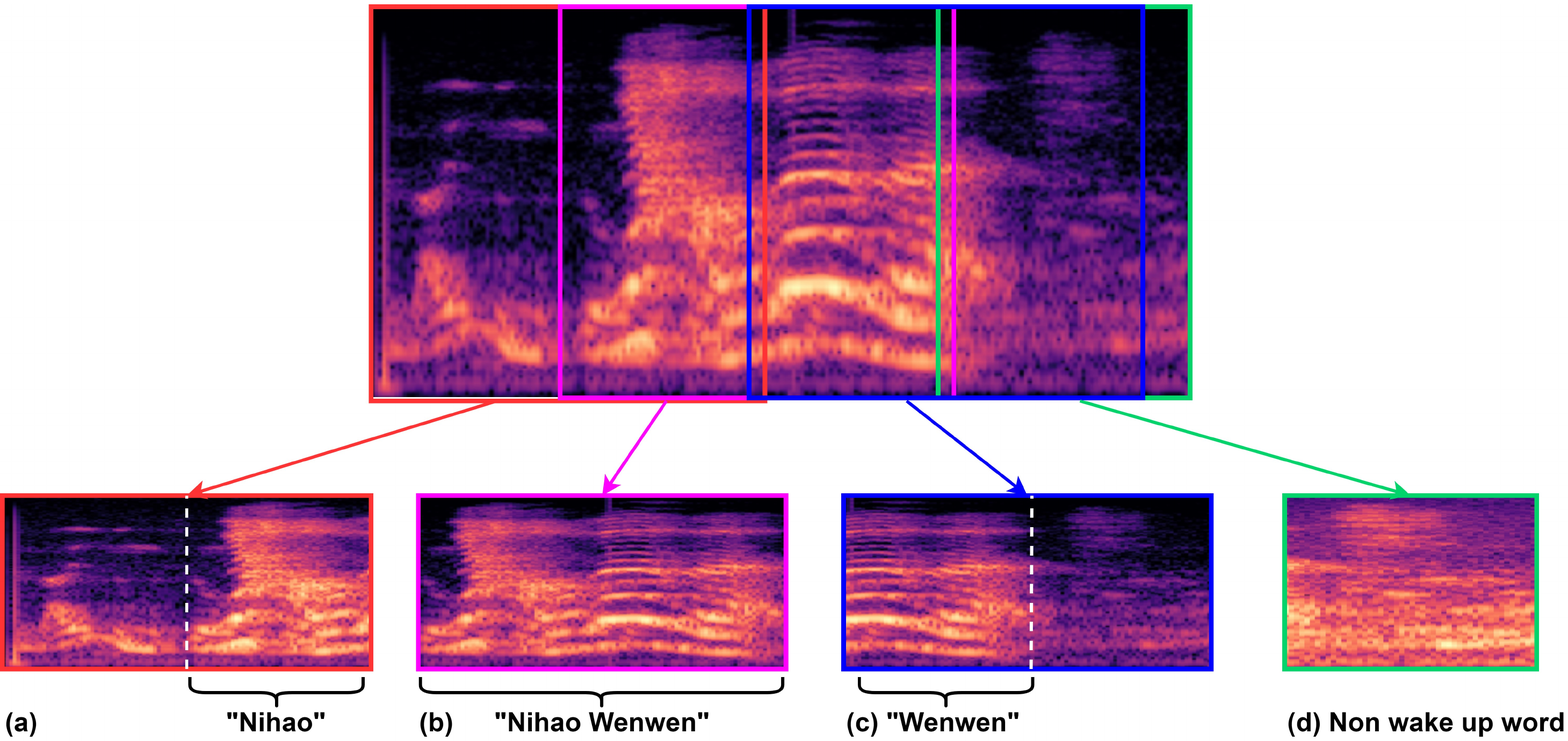}
    \caption{The spectrogram of "Nihao Wenwen" is divided into four sub-spectrograms, (a), (b), and (c) contain part or all of the wake word, and (d) does not contain any wake word.}
    \label{fig:yu2}
\end{figure}

This letter uses a sliding window with a length of $W$ and a step size of $s$ (where $s$ is the percentage of overlap between windows) on top of the spectrogram. The spectrogram is cut into many sub-spectrograms. The reason for using such a window is that it enables the classifier \textit{M1} to extract local patterns of wake words. The size of the window must be appropriate. A too-large window will lose the value of extracting local patterns, and a too-small window will make the model difficult to learn useful wake word features. Assuming that the wake word is "Nihao Wenwen" in Mandarin. As illustrated in Fig. \ref{fig:yu2}, sub-spectrograms (a), (b), and (c) are all wake word representations, and (d) does not contain any wake word. Meanwhile, the content in the sub-spectrogram is random, which may be "Nihao", "Wenwen", or "Nihao Wen" or "Hao Wenwen". To reduce the complexity of the training model, each sub-spectrogram inherits the label of the utterance where it lies.

\subsection{The classifiers M0 and M1}
The classifiers \textit{M0} and \textit{M1} respectively take the two types of spectrograms as inputs and finally output their scores. In the WWD system, SE-Res2Net is used as the classifier, which reduces the number of parameters.

\subsubsection{Res2Net Block}

\begin{figure}[h] \centering
    \includegraphics[width=0.4\textwidth, height=0.4\textwidth]{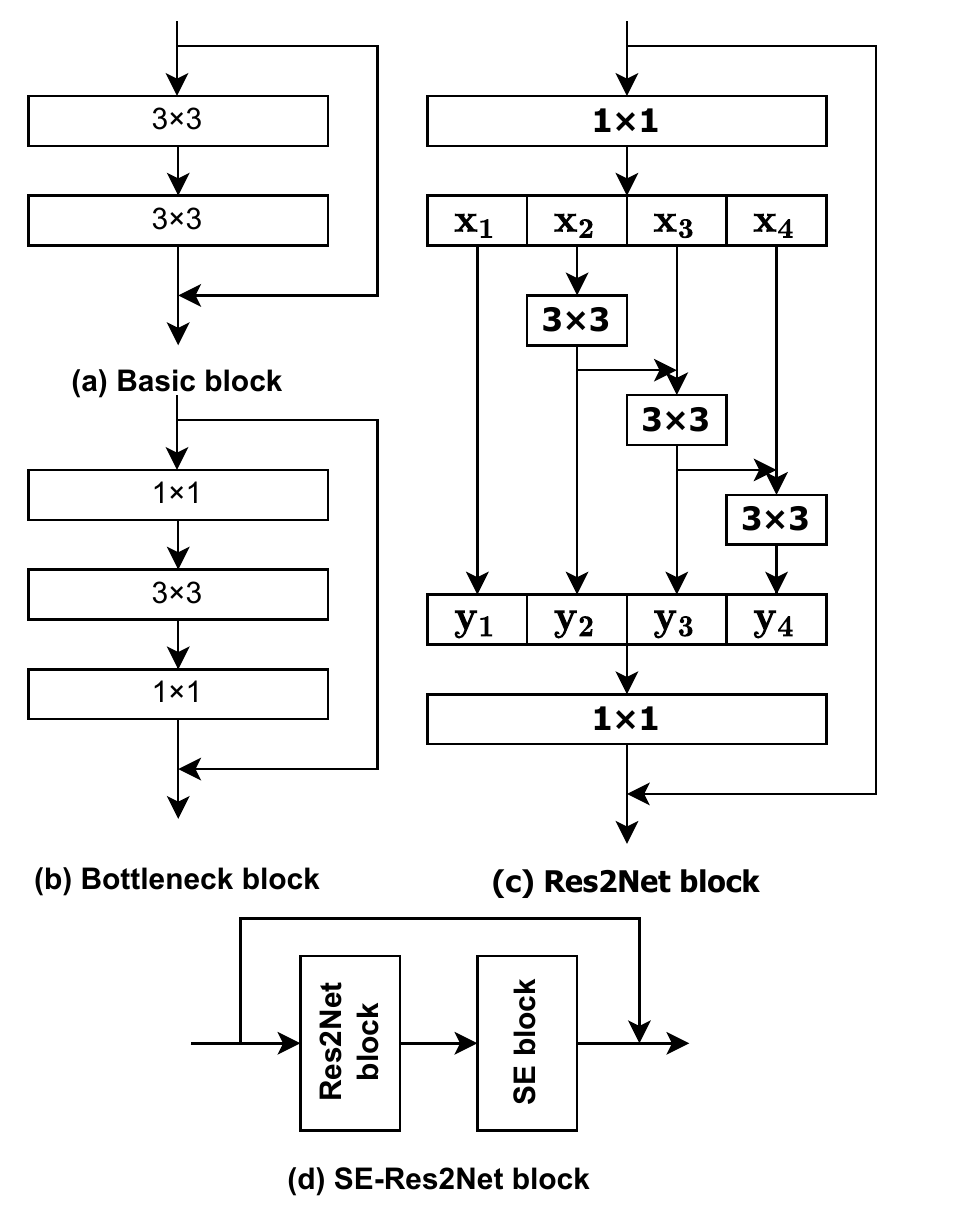}
    \caption{The basic block, bottleneck block, Res2Net block, and SE-Res2Net block (the scale dimension $s$=4; $y_1$, $y_2$, $y_3$, and $y_4$ represent the feature maps in a channel group).}
    \label{fig:yu3}
\end{figure}

A Res2Net block increases the number of receptive fields that are accessible in one layer to enhance the model's multi-scale representation. When multiple blocks are stacked, attributed to the combination effect, feature representations of various granularities can be obtained. Fig. \ref{fig:yu3} presents the comparison of the basic block, bottleneck block \cite{ResNet}, and Res2Net block. The Res2Net block is designed based on the bottleneck block. After processed by the first 1$\times$1 convolutional layer, the feature maps are split into \emph{s} subsets by the channel dimension, which are denoted as \{$x_{_1},x_{_2},...,x_{_s}$\}. Except for $x_{_1}$, each subset is fed into a 3$\times$3 convolution (denoted as $K_{_i}$). From $x_{_3}$, each $K_{_i-1}$ output is added with $x_{_i}$ before it passes through $K_{_i}$. This process is represented in Eq. \ref{conv}.

\begin{equation}
y_{i}=\begin{cases}
x_{i} &  i=1 \\
K_{i}\left ( x_{i} \right ) &  i=2 \\
K_{i}\left ( x_{i}+y_{i-1} \right ) &  2<i\leq s
\end{cases}
\label{conv}
\end{equation}

In Eq. \ref{conv}, \{$y_1,y_2,...,y_s$\} is the module's output, which represents the channel size of this residual block. It is concatenated and fed into the subsequent 1$\times$1 convolutional layer. In \cite{Res2Net}, \emph{s} is defined as the scale dimension.

\subsubsection{Integration with the squeeze-and-excitation (SE) block}
The SE block \cite{se_block} re-calibrates channel-wise feature responses. Through modeling the inter-dependencies between channels explicitly, different effect weights are assigned to the channels, thereby enabling the model to focus on the most relevant channel pattern of wake words. As shown in Fig. \ref{conv} (d), SE-Res2Net is formed by stacking Res2Net and SE blocks.

\begin{table*}[thb] 
    \large
    \centering
    \caption{The model of modified ResNet50, Res2Net50, and SE-Res2Net50. The repetition times of every block in a stage are defined outside the brackets, while the residual block type and the channel number are provided inside the brackets. ``2-d fc" represents a fully connected layer with two output units. The main difference from the original version is that the number of channels is reduced. ``\uppercase\expandafter{\romannumeral+ 1}" indicates that the channel number is changed to 1/4 of the original, and ``\uppercase\expandafter{\romannumeral+ 2}" indicates that the number of channels is reduced and the Conv5 layer is removed.}
    \label{tab:yu1}
	\resizebox{1\textwidth}{!}{
        \begin{tabular}[]{c | c | c | c | c | c | c}
        \hline 
        \multicolumn{1}{c|}{Stage} & ResNet50-\uppercase\expandafter{\romannumeral+ 1} & ResNet50-\uppercase\expandafter{\romannumeral+ 2} & Res2Net50-\uppercase\expandafter{\romannumeral+ 1} & Res2Net50-\uppercase\expandafter{\romannumeral+ 2} & SE-Res2Net50-\uppercase\expandafter{\romannumeral+ 1} & SE-Res2Net50-\uppercase\expandafter{\romannumeral+ 2} \\ 
        \hline
        \multicolumn{1}{c|}{\multirow{2}{*}{Conv1}} & \multicolumn{2}{c}{conv2d, 7$\times$7, 16, stride=2} & \multirow{2}{*}{{[}conv2d, 3$\times$3, 16, stride=1{]}$\times$3} & {[}conv2d, 3$\times$3, 16, stride=1{]}$\times$2 & \multirow{2}{*}{{[}conv2d, 3$\times$3, 16, stride=1{]}$\times$3} & {[}conv2d, 3$\times$3, 16, stride=1{]}$\times$2 \\
        \multicolumn{1}{c|}{} & \multicolumn{2}{c}{max pool, 3$\times$3, stride=2} & & conv2d, 3$\times$3, 16, stride=2 & & conv2d, 3$\times$3, 16, stride=2 \\ 
        \hline
        \multicolumn{1}{c|}{Conv2} & \multicolumn{1}{c|}{{[}Basic BLK, 4{]}$\times$3} & \multicolumn{1}{c|}{{[}Basic BLK, 4{]}$\times$3} & \multicolumn{1}{c|}{{[}Res2Net BLK, 4{]}$\times$3} & \multicolumn{1}{c|}{{[}Res2Net BLK, 4{]}$\times$3}  & \multicolumn{1}{c|}{{[}SE-Res2Net BLK, 4{]}$\times$3} & {[}SE-Res2Net BLK, 4{]}$\times$3 \\ 
        \hline
        \multicolumn{1}{c|}{Conv3} & \multicolumn{1}{c|}{{[}Basic BLK, 8{]}$\times$4}  & \multicolumn{1}{c|}{{[}Basic BLK, 8{]}$\times$4}  & \multicolumn{1}{c|}{{[}Res2Net BLK, 8{]}$\times$4} & \multicolumn{1}{c|}{{[}Res2Net BLK, 8{]}$\times$4}  & \multicolumn{1}{c|}{{[}SE-Res2Net BLK, 8{]}$\times$4} & {[}SE-Res2Net BLK, 8{]}$\times$4 \\ 
        \hline
        \multicolumn{1}{c|}{Conv4} & \multicolumn{1}{c|}{{[}Basic BLK, 16{]}$\times$6} & \multicolumn{1}{c|}{{[}Basic BLK, 16{]}$\times$6} & \multicolumn{1}{c|}{{[}Res2Net BLK, 16{]}$\times$6} & \multicolumn{1}{c|}{{[}Res2Net BLK, 16{]}$\times$6} & \multicolumn{1}{c|}{{[}SE-Res2Net BLK, 16{]}$\times$6} & {[}SE-Res2Net BLK, 16{]}$\times$6 \\ 
        \hline
        \multicolumn{1}{c|}{Conv5} & \multicolumn{1}{c|}{{[}Basic BLK, 32{]}$\times$3} & \multicolumn{1}{c|}{-} & \multicolumn{1}{c|}{{[}Res2Net BLK, 32{]}$\times$3} & \multicolumn{1}{c|}{-} & \multicolumn{1}{c|}{{[}SE-Res2Net BLK, 32{]}$\times$3} & - \\ 
        \hline
        \multicolumn{7}{c}{global average pool, 2-d fc, softmax} \\
        \hline
        \end{tabular}
        }
\end{table*}

\subsubsection{Streaming inference}
When processing streaming audio, \textit{M1} first detects each sub-spectrogram of the audio. When the score of the sub-spectrogram exceeds the threshold $\gamma_{1}$, it is considered as a trigger point. Once there are at least two consecutive trigger points, there may be wake words in this part of the audio. The maximum score among consecutive trigger points is denoted as $y_{M1}$. Meanwhile, \textit{M0} intervenes to detect the spectrogram between the first trigger point and the last trigger point to obtain $y_{M0}$. The final score $y_{f}$ is the average of $y_{M0}$ and $y_{M1}$. If $y_{f}$ is larger than $\gamma_{1}$, it is said that the wake word is triggered. Once a wake word is triggered, the hits in the next second are neglected. $\gamma_{1}$ $\in (0,1)$ is a confidence threshold, and it is adjusted on the development set.

\section{EXPERIMENTS}
\subsection{The dataset}
This letter uses the Mobvoi (SLR87) dataset \cite{pure_nn5} containing 2 wake words: "Hi Xiaowen" and "Nihao Wenwen". Both wake words are recorded by all the speakers. The dataset includes 144 hours of training data (174,592 samples in total, where 43,625 are positive ones) and 74 hours of test data (73,459 samples in total, where 21,282 are positive). Here, separate models are trained for each wake word, and this is considered as a binary classification problem. When considering one of the two wake words, the other is considered negative. 

\subsection{Experimental Settings}
In all experiments, for every utterance, spectrograms are calculated by using 1024 discrete Fourier transform points, a 256-band Mel scale, as well as a hop size of 160. Since the length of both wake words is between 30 frames and 200 frames, this letter uses bilinear interpolation to adjust all spectrograms to 200 frames to train \textit{M0}. For \textit{M1}, different sliding windows are used for different wake words. The window length $w$ of "Hi Xiaowen" is 75 frames, and the step size $s$ is 0.3; the $w$ of "Nihao Wenwen" is 100 frames, and the $s$ = 0.3. The network architecture of the classifier used in this letter is presented in Tabel \ref{tab:yu1}. Besides, Adam optimization \cite{Adam} is employed with an initial learning rate of 0.0002. In the first 5 training epochs, the SpecAugment strategy \cite{specaugment} is used. For each training utterance, 0 - 30 consecutive frames are randomly selected, and all their mel-filter banks are set to 0 for time masking. For frequency masking, 0 - 20 consecutive dimensions of the 256 mel-filter banks are randomly selected, and their values are set to 0 for all frames. Note that in a training mini-batch, only one-third of the utterances receive the time masking, the other one-third receive the frequency masking, and the remaining utterances receive both maskings. The training process does not stop when the number of epochs is smaller than 20 or the learning rate is more than 0.01.

\subsection{Results}

\subsubsection{Effect of the proposed system}

To compare the SE-Res2Net block with the Res2Net block and the basic block for our WWD system, the Res2Net50 and ResNet50 shown in Table \ref{tab:yu1} are used as classifier models to train the comparison system. Also, the parameters of Res2Net50 and ResNet50 are reduced in the same way. Except for the different models, the remaining settings are the same.

\begin{figure}[thb] \centering
    \includegraphics[width=0.45\textwidth, height=0.45\textwidth]{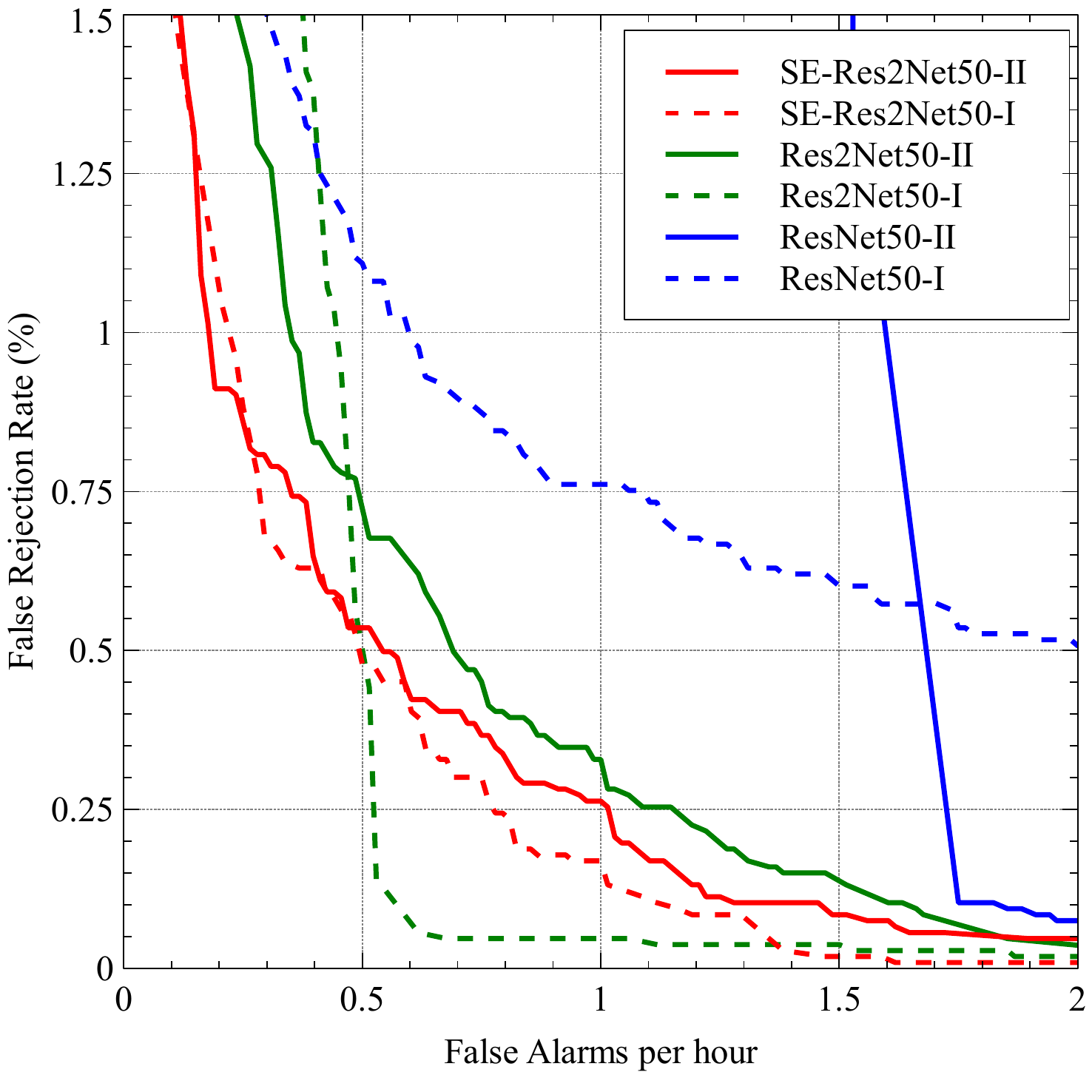}
    \caption{The DET curves for ``Hi Xiaowen"} \label{fig:yu4}
\end{figure}

\begin{figure}[thb] \centering
    \includegraphics[width=0.45\textwidth, height=0.45\textwidth]{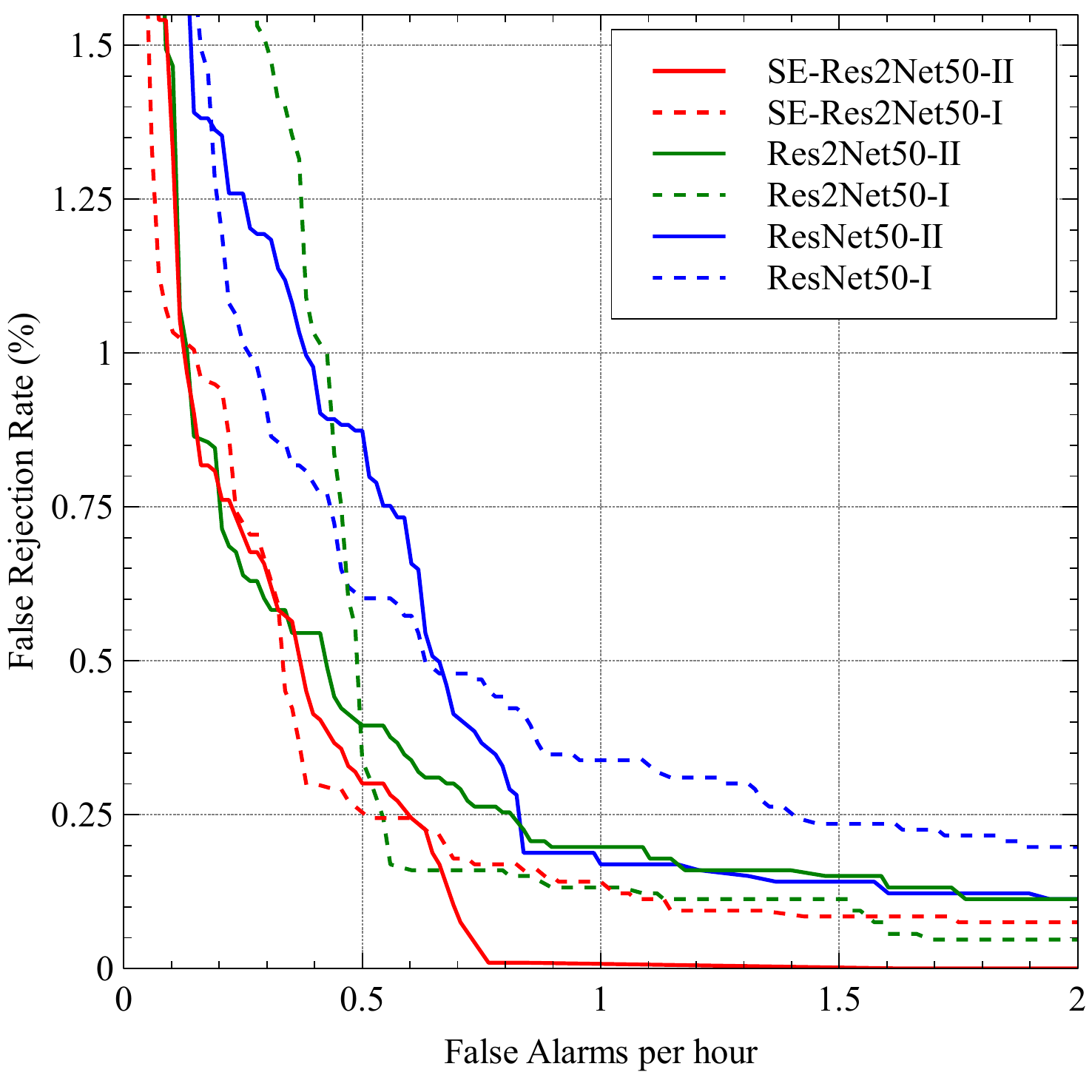}
    \caption{The DET curves for ``Nihao Wenwen"} \label{fig:yu5}
\end{figure}

The DET curves of the wake words are presented in Figs. \ref{fig:yu4} and \ref{fig:yu5}. Comparing ResNet50 and Res2Net50, it is indicated that Res2Net50 performs better than ResNet50 in all conditions. This shows that the Res2Net block improves the detection ability of wake words through multi-granularity features. After taking
SE-Res2Net50 into comparison, it significantly outperforms both Res2Net50 and ResNet50. More specifically, at FAH = 0.5, SE-Res2Net50-\uppercase\expandafter{\romannumeral+ 2} outperforms ResNet50-\uppercase\expandafter{\romannumeral+ 2} and Res2Net50-\uppercase\expandafter{\romannumeral+ 2} with an FRR reduction of 86.6\% and 31.2\% on "Hi Xiaowen", and 65.5\% and 23.1\% on "Nihao Wenwen", respectively. Similar gains can be also obtained at the other set of experiments. Furthermore, it is found that the DET curve of "Nihao Wenwen" is better than that of "Hi Xiaowen". This may be related to the fact that "Nihao Wenwen" has more syllables (4 rather than 3) and can be identified from other non-keyword audio more easily.

\subsubsection{Comparison with prior works}
Then, our proposed WWD system is compared with three recent baselines on the Mobvoi (SLR87) dataset \cite{pure_nn5, pure_nn4, pure_nn15}. \cite{pure_nn4} used an additional HMM-TDNN acoustic model during training. \cite{pure_nn5,pure_nn15} are hybrid HMM/DNN system with alignment-free LF-MMI loss, without auxiliary system. The results are listed in Table \ref{tab:yu2}. Under FAH=0.5, our system achieves a 12-50\% lower FRR than the transformer-based system \cite{pure_nn15}. Meanwhile, our system achieves a similar performance with the HMM-DNN-based system \cite{pure_nn5} with only half model parameters and without any additional network.

\begin{table}[h] \scriptsize
    \caption{Comparison with other WWD systems}
    \label{tab:yu2}
    \resizebox{0.48\textwidth}{!}{
    \begin{tabular}{l c c c }
    \hline
    \multirow{2}{*}{\textit{Mobvoi (SLR87)}} & \multirow{2}{*}{\#Params} & \multicolumn{2}{c}{FRR(\%) at FAH=0.5} \\
    \cline{3-4}
     & & Hi xiaowen & Nihao Wenwen \\
    \hline
    Negative data mining algorithm system \cite{pure_nn4} & N/A & 1.3 & 1.0 \\
    \hline
     Alignment-Free Lattice-Free MMI system \cite{pure_nn5} & 120K & 0.4 & 0.5 \\
    \hline
    Streaming transformers system \cite{pure_nn15} & 57k & 0.6 & 0.6 \\
    \hline
    SE-Res2Net50-\uppercase\expandafter{\romannumeral+ 1} based System & 128K & 0.47 & 0.25 \\
    \hline
    SE-Res2Net50-\uppercase\expandafter{\romannumeral+ 2} based System & 52K & 0.53 & 0.30 \\
    \hline
    \end{tabular}}
\end{table}

\section{SUMMARY}

In this letter, a novel CNN-based WWD system is proposed. The proposed system does not rely on additional networks for forced alignment but trains two classifiers through two types of feature processing. One classifier detects the feature slice segments of wake words to roughly locate their positions, and the other classifier further detects the speech segments after localization. Meanwhile, Res2Net, an improved ResNet network, is used as the classifier model. The Res2Net block is used to expand the possible receptive fields and improve the detection ability of the classifier. The effectiveness of our proposed system is verified on two commercial wake words. Compared with prior systems, our system can achieve better performance with fewer model parameters.

\bibliographystyle{IEEEtran}
\bibliography{cite}

\begin{thebibliography}{10}
\providecommand{\url}[1]{#1}
\csname url@samestyle\endcsname
\providecommand{\newblock}{\relax}
\providecommand{\bibinfo}[2]{#2}
\providecommand{\BIBentrySTDinterwordspacing}{\spaceskip=0pt\relax}
\providecommand{\BIBentryALTinterwordstretchfactor}{4}
\providecommand{\BIBentryALTinterwordspacing}{\spaceskip=\fontdimen2\font plus
\BIBentryALTinterwordstretchfactor\fontdimen3\font minus
  \fontdimen4\font\relax}
\providecommand{\BIBforeignlanguage}[2]{{%
\expandafter\ifx\csname l@#1\endcsname\relax
\typeout{** WARNING: IEEEtran.bst: No hyphenation pattern has been}%
\typeout{** loaded for the language `#1'. Using the pattern for}%
\typeout{** the default language instead.}%
\else
\language=\csname l@#1\endcsname
\fi
#2}}
\providecommand{\BIBdecl}{\relax}
\BIBdecl

\bibitem{keyword_search}
J.~Trmal, M.~Wiesner, V.~Peddinti, X.~Zhang, P.~Ghahremani, Y.~Wang,
  V.~Manohar, H.~Xu, D.~Povey, and S.~Khudanpur, ``The kaldi openkws system:
  Improving low resource keyword search.'' in \emph{Interspeech}, 2017, pp.
  3597--3601.

\bibitem{lvcsr1}
M.~Saraclar and R.~Sproat, ``Lattice-based search for spoken utterance
  retrieval,'' in \emph{Proceedings of the Human Language Technology Conference
  of the North American Chapter of the Association for Computational
  Linguistics: HLT-NAACL 2004}, 2004, pp. 129--136.

\bibitem{lvcsr2}
D.~Can and M.~Saraclar, ``Lattice indexing for spoken term detection,''
  \emph{IEEE Transactions on Audio, Speech, and Language Processing}, vol.~19,
  no.~8, pp. 2338--2347, 2011.

\bibitem{lvcsr3}
J.~Mamou, B.~Ramabhadran, and O.~Siohan, ``Vocabulary independent spoken term
  detection,'' in \emph{Proceedings of the 30th annual international ACM SIGIR
  conference on Research and development in information retrieval}, 2007, pp.
  615--622.

\bibitem{hmm1}
R.~C. Rose and D.~B. Paul, ``A hidden markov model based keyword recognition
  system,'' in \emph{International Conference on Acoustics, Speech, and Signal
  Processing}.\hskip 1em plus 0.5em minus 0.4em\relax IEEE, 1990, pp. 129--132.

\bibitem{hmm2}
I.~Sz{\"o}ke, P.~Schwarz, P.~Mat{\v{e}}jka, L.~Burget, M.~Karafi{\'a}t, and
  J.~{\v{C}}ernock{\`y}, ``Phoneme based acoustics keyword spotting in informal
  continuous speech,'' in \emph{International Conference on Text, Speech and
  Dialogue}.\hskip 1em plus 0.5em minus 0.4em\relax Springer, 2005, pp.
  302--309.

\bibitem{hmm3}
M.-C. Silaghi, ``Spotting subsequences matching an hmm using the average
  observation probability criteria with application to keyword spotting,'' in
  \emph{AAAI}, 2005, pp. 1118--1123.

\bibitem{pure_nn1}
G.~Chen, C.~Parada, and G.~Heigold, ``Small-footprint keyword spotting using
  deep neural networks,'' in \emph{2014 IEEE International Conference on
  Acoustics, Speech and Signal Processing (ICASSP)}.\hskip 1em plus 0.5em minus
  0.4em\relax IEEE, 2014, pp. 4087--4091.

\bibitem{pure_nn2}
T.~Sainath and C.~Parada, ``Convolutional neural networks for small-footprint
  keyword spotting,'' 2015.

\bibitem{pure_nn3}
Y.~He, R.~Prabhavalkar, K.~Rao, W.~Li, A.~Bakhtin, and I.~McGraw, ``Streaming
  small-footprint keyword spotting using sequence-to-sequence models,'' in
  \emph{2017 IEEE Automatic Speech Recognition and Understanding Workshop
  (ASRU)}.\hskip 1em plus 0.5em minus 0.4em\relax IEEE, 2017, pp. 474--481.

\bibitem{hmm_dnn4}
S.~Panchapagesan, M.~Sun, A.~Khare, S.~Matsoukas, A.~Mandal, B.~Hoffmeister,
  and S.~Vitaladevuni, ``Multi-task learning and weighted cross-entropy for
  dnn-based keyword spotting.'' in \emph{Interspeech}, vol.~9, 2016, pp.
  760--764.

\bibitem{hmm_dnn5}
M.~Sun, D.~Snyder, Y.~Gao, V.~K. Nagaraja, M.~Rodehorst, S.~Panchapagesan,
  N.~Strom, S.~Matsoukas, and S.~Vitaladevuni, ``Compressed time delay neural
  network for small-footprint keyword spotting.'' in \emph{Interspeech}, 2017,
  pp. 3607--3611.

\bibitem{hmm_dnn6}
M.~Wu, S.~Panchapagesan, M.~Sun, J.~Gu, R.~Thomas, S.~N.~P. Vitaladevuni,
  B.~Hoffmeister, and A.~Mandal, ``Monophone-based background modeling for
  two-stage on-device wake word detection,'' in \emph{2018 IEEE International
  Conference on Acoustics, Speech and Signal Processing (ICASSP)}.\hskip 1em
  plus 0.5em minus 0.4em\relax IEEE, 2018, pp. 5494--5498.

\bibitem{pure_nn6}
S.~Myer and V.~S. Tomar, ``Efficient keyword spotting using time delay neural
  networks,'' \emph{arXiv preprint arXiv:1807.04353}, 2018.

\bibitem{pure_nn7}
A.~Coucke, M.~Chlieh, T.~Gisselbrecht, D.~Leroy, M.~Poumeyrol, and T.~Lavril,
  ``Efficient keyword spotting using dilated convolutions and gating,'' in
  \emph{ICASSP 2019-2019 IEEE International Conference on Acoustics, Speech and
  Signal Processing (ICASSP)}.\hskip 1em plus 0.5em minus 0.4em\relax IEEE,
  2019, pp. 6351--6355.

\bibitem{pure_nn10}
T.~N. Sainath, O.~Vinyals, A.~Senior, and H.~Sak, ``Convolutional, long
  short-term memory, fully connected deep neural networks,'' in \emph{2015 IEEE
  international conference on acoustics, speech and signal processing
  (ICASSP)}.\hskip 1em plus 0.5em minus 0.4em\relax IEEE, 2015, pp. 4580--4584.

\bibitem{pure_nn12}
M.~Sun, A.~Raju, G.~Tucker, S.~Panchapagesan, G.~Fu, A.~Mandal, S.~Matsoukas,
  N.~Strom, and S.~Vitaladevuni, ``Max-pooling loss training of long short-term
  memory networks for small-footprint keyword spotting,'' in \emph{2016 IEEE
  Spoken Language Technology Workshop (SLT)}.\hskip 1em plus 0.5em minus
  0.4em\relax IEEE, 2016, pp. 474--480.

\bibitem{hochreiter1997long}
S.~Hochreiter and J.~Schmidhuber, ``Long short-term memory,'' \emph{Neural
  computation}, vol.~9, no.~8, pp. 1735--1780, 1997.

\bibitem{cho2014learning}
K.~Cho, B.~Van~Merri{\"e}nboer, C.~Gulcehre, D.~Bahdanau, F.~Bougares,
  H.~Schwenk, and Y.~Bengio, ``Learning phrase representations using rnn
  encoder-decoder for statistical machine translation,'' \emph{arXiv preprint
  arXiv:1406.1078}, 2014.

\bibitem{pure_nn16}
Y.~Wang \emph{et~al.}, ``Wake word detection and its applications,'' Ph.D.
  dissertation, Johns Hopkins University, 2021.

\bibitem{arik2017convolutional}
S.~O. Arik, M.~Kliegl, R.~Child, J.~Hestness, A.~Gibiansky, C.~Fougner,
  R.~Prenger, and A.~Coates, ``Convolutional recurrent neural networks for
  small-footprint keyword spotting,'' \emph{arXiv preprint arXiv:1703.05390},
  2017.

\bibitem{bai2019time}
Y.~Bai, J.~Yi, J.~Tao, Z.~Wen, Z.~Tian, C.~Zhao, and C.~Fan, ``A time delay
  neural network with shared weight self-attention for small-footprint keyword
  spotting.'' in \emph{INTERSPEECH}, 2019, pp. 2190--2194.

\bibitem{res2net1}
S.-H. Gao, M.-M. Cheng, K.~Zhao, X.-Y. Zhang, M.-H. Yang, and P.~Torr, ``A new
  multi-scale backbone architecture,'' \emph{IEEE transactions on pattern
  analysis and machine intelligence}, vol.~43, no.~2, pp. 652--662, 2019.

\bibitem{sharan2019time}
R.~V. Sharan and T.~J. Moir, ``Time-frequency image resizing using
  interpolation for acoustic event recognition with convolutional neural
  networks,'' in \emph{2019 IEEE International Conference on Signals and
  Systems (ICSigSys)}.\hskip 1em plus 0.5em minus 0.4em\relax IEEE, 2019, pp.
  8--11.

\bibitem{ResNet}
K.~He, X.~Zhang, S.~Ren, and J.~Sun, ``Deep residual learning for image
  recognition,'' in \emph{Proceedings of the IEEE conference on computer vision
  and pattern recognition}, 2016, pp. 770--778.

\bibitem{se_block}
J.~Hu, L.~Shen, and G.~Sun, ``Squeeze-and-excitation networks,'' in
  \emph{Proceedings of the IEEE conference on computer vision and pattern
  recognition}, 2018, pp. 7132--7141.

\bibitem{pure_nn5}
Y.~Wang, H.~Lv, D.~Povey, L.~Xie, and S.~Khudanpur, ``Wake word detection with
  streaming transformers,'' in \emph{ICASSP 2021-2021 IEEE International
  Conference on Acoustics, Speech and Signal Processing (ICASSP)}.\hskip 1em
  plus 0.5em minus 0.4em\relax IEEE, 2021, pp. 5864--5868.

\bibitem{Adam}
D.~P. Kingma and J.~Ba, ``Adam: A method for stochastic optimization,''
  \emph{arXiv preprint arXiv:1412.6980}, 2014.

\bibitem{specaugment}
D.~S. Park, W.~Chan, Y.~Zhang, C.-C. Chiu, B.~Zoph, E.~D. Cubuk, and Q.~V. Le,
  ``Specaugment: A simple data augmentation method for automatic speech
  recognition,'' \emph{arXiv preprint arXiv:1904.08779}, 2019.

\bibitem{pure_nn4}
J.~Hou, Y.~Shi, M.~Ostendorf, M.-Y. Hwang, and L.~Xie, ``Mining effective
  negative training samples for keyword spotting,'' in \emph{ICASSP 2020-2020
  IEEE International Conference on Acoustics, Speech and Signal Processing
  (ICASSP)}.\hskip 1em plus 0.5em minus 0.4em\relax IEEE, 2020, pp. 7444--7448.

\bibitem{pure_nn15}
Y.~Wang, H.~Lv, D.~Povey, L.~Xie, and S.~Khudanpur, ``Wake word detection with
  alignment-free lattice-free mmi,'' \emph{arXiv preprint arXiv:2005.08347},
  2020.

\end{thebibliography}

\end{document}